\documentclass[3p, review,12pt]{elsarticle}

\usepackage[colorinlistoftodos]{todonotes}
\usepackage{placeins}
\usepackage{graphicx}
\usepackage{ecrc}

\volume{00}
\firstpage{1}
\journalname{Journal of Membrane Science}
\runauth{K. Baitalow}
\jid{J. Memb. Sci.}
\jnltitlelogo{J. Memb. Sci.}

\usepackage{siunitx} 
\usepackage{verbatim} 
\usepackage{subfig} 
\usepackage{microtype} 

\usepackage{amssymb}
\usepackage[version=4]{mhchem}
\usepackage{overpic}
\usepackage{graphics}
\usepackage{multirow}
\usepackage{rotating}
\usepackage{textcomp}
\usepackage{float}
\usepackage{color}

\begin{document}

\begin{frontmatter}

\title{A mini-module with built-in spacers for high-throughput ultrafiltration }

\author[CVT]{Kristina Baitalow}
\author[CVT,DWI]{Denis Wypysek}
\author[Sartorius]{Martin Leuthold}
\author[Sartorius]{Stefan Weisshaar}
\author[CVT,DWI]{Jonas L\"olsberg}
\author[CVT,DWI]{Matthias Wessling\corref{mycorrespondingauthor}}

\cortext[mycorrespondingauthor]{Corresponding author: manuscripts.cvt@avt.rwth-aachen.de}

\address[CVT]{RWTH Aachen University, Chemical Process Engineering, Forckenbeckstrasse 51, 52074 Aachen, Germany}
\address[DWI]{DWI-Leibniz - Institute for Interactive Materials, Forckenbeckstrasse 50, 52074 Aachen, Germany}
\address[Sartorius]{Sartorius Stedim Biotech GmbH, August-Spindler-Str. 11, 37079 G\"ottingen, Germany}

\begin{abstract}
Ultrafiltration membrane modules suffer from a permeate flow decrease arising during filtration and caused by concentration polarization and fouling in, e.g., fermentation broth purification. Such performance losses are frequently mitigated by manipulating the hydrodynamic conditions at the membrane-fluid interface using, e.g., mesh spacers acting as static mixers. This additional element increases manufacturing complexity while improving mass transport in general, yet accepting their known disadvantages such as less transport in dead zones. However, the shape of such spacers is limited to the design of commercially available spacer geometries. Here, we present a methodology to design an industrially relevant mini-module with an optimized built-in 3D spacer structure in a flat-sheet ultrafiltration membrane module to eliminate the spacer as a separate part. Therefore, the built-in structures have been conceptually implemented through an in-silico design in compliance with the specifications for an injection molding process. Ten built-in structures were investigated in a digital twin of the mini-module by 3D-CFD simulations to select two options, which were then compared to the empty feed channel regarding mass transfer. Subsequently, the simulated flux increase was experimentally verified during bovine serum albumin (BSA) filtration. The new built-in sinusoidal corrugation outperforms conventional mesh spacer inlays by up to \SI{30}{\percent} higher permeation rates. The origin of these improvements is correlated to the flow characteristics inside the mini-module as visualized online and in-situ by low-field and high-field magnetic resonance imaging velocimetry (flow-MRI) during pure water permeation. 
\end{abstract}


\end{frontmatter}
\section{Introduction}
 
Ultrafiltration processes play an essential role in the biotechnological and pharmaceutical industries~\cite{Belfort1994, VanReis2007, Charcosset2012}. The key function of ultrafiltration (UF) membranes is removing harmful matter such as bacteria and viruses, concentrating a desired product such as proteins and enzymes, or separate valuable products to increase purity~\cite{ALAANI.2020, Lutz2015}. 

Particulate and colloidal matter is especially omnipresent in biotechnological processes. It tends to accumulate at the membrane surface, establishing a concentration polarization layer, and forming a complex deposit film on top of the membrane \cite{Belfort1989}. This fouling leads to a decrease in the solvent flux through the membranes \cite{vandenBerg1990}. Matthiasson and Sivik have already stated in the 1980s that concentration polarization occurring in membrane filtration processes leads to mass transfer inhibition \cite{Matthiasson1980}. Various issues have been taken to address this issue, including static mixers inserted into membrane modules to increase mixing and, subsequently, mass transfer \cite{Bhattacharjee2020}. 


Obstacles in the flow field reduce the critical Reynolds number to initiate oscillations, which highly depend on the fluid velocity and spacer geometry \cite{Geraldes2002, Rodrigues2012, Qamar2019}. Static mixers function as such obstacles disrupting the laminar flow profiles in the boundary layer, causing vortices in the flow field \cite{Schwinge2002}, effectively reducing concentration polarization at low Reynolds numbers. Qamar et al. approached the gap of a comprehensive study on the transition from steady to unsteady flow~\cite{Qamar2019}. Obstacle-induced mixing also leads to higher velocities and shear rates at the membrane surface. The locally increased shear rate has, in turn, a positive effect on reducing fouling~\cite{DaCosta1993, Picioreanu2009} induced by biological or colloidal matter. However, there is a maximum shear rate, above which biological matter is likely to be destroyed and can favor biofouling~\cite{Qamar2019, Picioreanu2009}. Computational fluid dynamics (CFD) simulations of mesh-type spacers show a better mixing (high shear rate, low drag coefficient) at the expense of an increase in axial pressure drop along the feed channel~\cite{Karode2001, Li2002, Picioreanu2009, Qamar2019}. This leads to a constant trade-off between flux increase and the correlated rise in energy input due to higher pressure loss~\cite{DaCosta1991, Kavianipour2017}. 

Studies on net- or diamond-shaped spacers and meshes are available in large numbers and are discussed in several reviews \cite{Kavianipour2017, Abid2017, Lee2016, FimbresWeihs2010, Koo2021}. 
The literature on experimental and simulation-based investigations of mesh spacers proves reduced biofouling and prolonged stable membrane-module performance \cite{DaCosta1991, Kodym2011, Rivera2017, Bucs2015, Thiess2017, Johannink2015, Picioreanu2009}. On a simple basis, the literature describes such mesh spacers or static mixers using 2D structures like triangles, squares, or circles and focuses on their position and dimension \cite{Schwinge2002, Ahmad2005, Cao2001}. Such studies support the understanding of flow-dynamic correlations as a pre-step on the way to new innovative design solutions, yet simplifications sometimes are too rigorous and correct representation of the genuine spacers details is questionable ~\cite{Picioreanu2009}.

One of the less considered downsides of mesh-type spacers is that they require several assembling steps during in-line module assembly. Here, incorporation into the module housing is proposed with an anticipated single-step production by injection molding, accelerating the production process and reducing operating costs. Before establishing such a production process, we propose to follow a methodology of in-silico design combined with 3D-printing-based rapid prototyping. Removing such mesh spacers from a module architecture has been addressed by incorporating spacers as turbulence promoters into the membrane as so-called membrane-con-spacer geometries for various membrane-surface structures~\cite{Balster2010, Racz1986, Mazinani.2019, Lee2013, Zhou2021}. A patterned membrane surface acts as a spacer structure and induces vortices~\cite{Lee2013}. Typical geometries include flat, line and groove patterns, pyramids, rectangular and circular pillars over the whole channel width , or individual elevations. Detailed studies on sinusoidal membrane surfaces investigate the effect of sinusoidal structures and the resulting hydrodynamics on fouling resistance~\cite{Mazinani.2019, Lee2013}.

Implementation of commercial mesh spacers used as built-in spacer structures exists only for simple geometries such as squares~\cite{LiraTeco2016} or ribs~\cite{Completo2016}. Flow-aligned, bar-like filaments in a feed channel showed that flow is affected more by transverse than by longitudinal alignments \cite{Santos2007}. A similar structure, a ladder-type spacer within a frame fabricated by 3D printing, was studied in ultrafiltration experiments \cite{Liu2013} and through experiments and simulations of nanofiltration and reverse osmosis (RO)~processes\cite{Geraldes2002}. Both studies showed that an increasing inter-filament distance reduces the critical Reynolds number for the transition from laminar to turbulent flow. Here, the spacer elements direct the fluid flow to the upper and lower parts of the membrane module but contain undercuts, which is not suitable for injection molding. 
Other experiments show positive effects only in the entrance region for mesh-type spacers and embedded ribbon-type spacers \cite{Rodrigues2012}. Another built-in spacer type suitable for ultrafiltration is a zigzag spacer, which outperforms mesh-type spacers in terms of pressure drop \cite{Schwinge2000}. Staggered herringbone structures are a good example for built-in spacer structures but are suitable primarily for microfluidic applications \cite{Jung2019}. 
Shrivastava et al. \cite{Shrivastava2008} examined asymmetric herringbone, helix, and ladder-type structures, intending to provide a guide for developing reasonable spacers. Positive effects in the mass transfer were highest for helices, followed by herringbones and ladder-type spacers \cite{Shrivastava2008}. 

While much work tries to understand flow distribution and hydrodynamics, some studies also address the corresponding simultaneous mass transfer phenomena quantitatively \cite{Schwinge2002b, DaCosta1991, Koutsou2004, Schock1987,vanderWaal1989, Completo2016, Picioreanu2009}. At the membrane, high shear stress and the formation of recirculation zones and vortices are desired, enhancing the intermixing of material adjacent to the membrane \cite{Schwinge2002b}. Mass transfer enhancement is accompanied by an increase in axial pressure drop \cite{Schock1987}. This results in an increase in mass transfer coefficients with increasing shear rates \cite{Schwinge2002b, Koutsou2004}. Mass transfer investigations were considered for simple geometries only \cite{Schock1987, Koutsou2004}, but recently our group has addressed membranes with spacers \cite{BALSTER2006, Balster2010,Wiese.2018} and 3D-printed novel spacer and mixer geometries \cite{FRITZMANN.2013,FRITZMANN.2014, ARMBRUSTER.2018}. A general overview is given by the Virtual Issue of the Journal \cite{Popovic2015}.

Some studies investigate 3D-printing-aided membrane modules due to the great potential of 3D printing for fast and cheap module improvement \cite{Femmer2016f,Koo2021}. Module design and material selection must be considered in 3D printing, but research on 3D printing of membrane-process-related questions progresses rather slow~\cite{Lee2016, Koo2021}. 

So far, fundamental research was done focused on fluid flow simulations, simple 2D structures, or inlay mesh-type spacers to better understand correlations between feed channel obstacles and fouling behavior. 
3D simulations are preferable to 2D simulations because the latter underestimate the parameters examined and the three-dimensional effects of intermixing~\cite{Kodym2011, FimbresWeihs2010, Picioreanu2009}. 

In this study, we built on all this preliminary work and used CFD as a widely established tool for evaluating flow and concentration behavior in a digital twin. The novelty of this work consists in the methodology to (a) combine CFD simulations within a digital twin, (b) protein fouling experiments with a 3D printed copy of the digital twin, and (c) MRI investigations to prove the presence of flow patterns as predicted by the CFD simulations. We performed 3D-CFD simulations to compare ten potential built-in structures regarding shear rate distribution and pressure loss. Subsequently, the simulations of the two structures with the most promising shear rate-to-pressure loss ratio were extended by mass transfer correlations. To verify the simulated permeation increase, these two structures were implemented in a high-throughput membrane module fabricated by 3D printing for fast prototyping and tested in bovine serum albumin (BSA) filtration experiments. The positive effect of the chosen structure in the final product was correlated to hydrodynamic effects visualized by online and in-situ low-field (LF) as well as high-field (HF) flow-MRI measurements. We complement CFD simulations with both fouling experiments and flow-MRI measurements for a qualitative comparison and to better understand prevailing hydrodynamic phenomena.

\section{Materials and methods}
\subsection{Additive manufactured membrane modules}
\begin{figure}[H]
    \centering
    \includegraphics[width=0.45\textwidth]{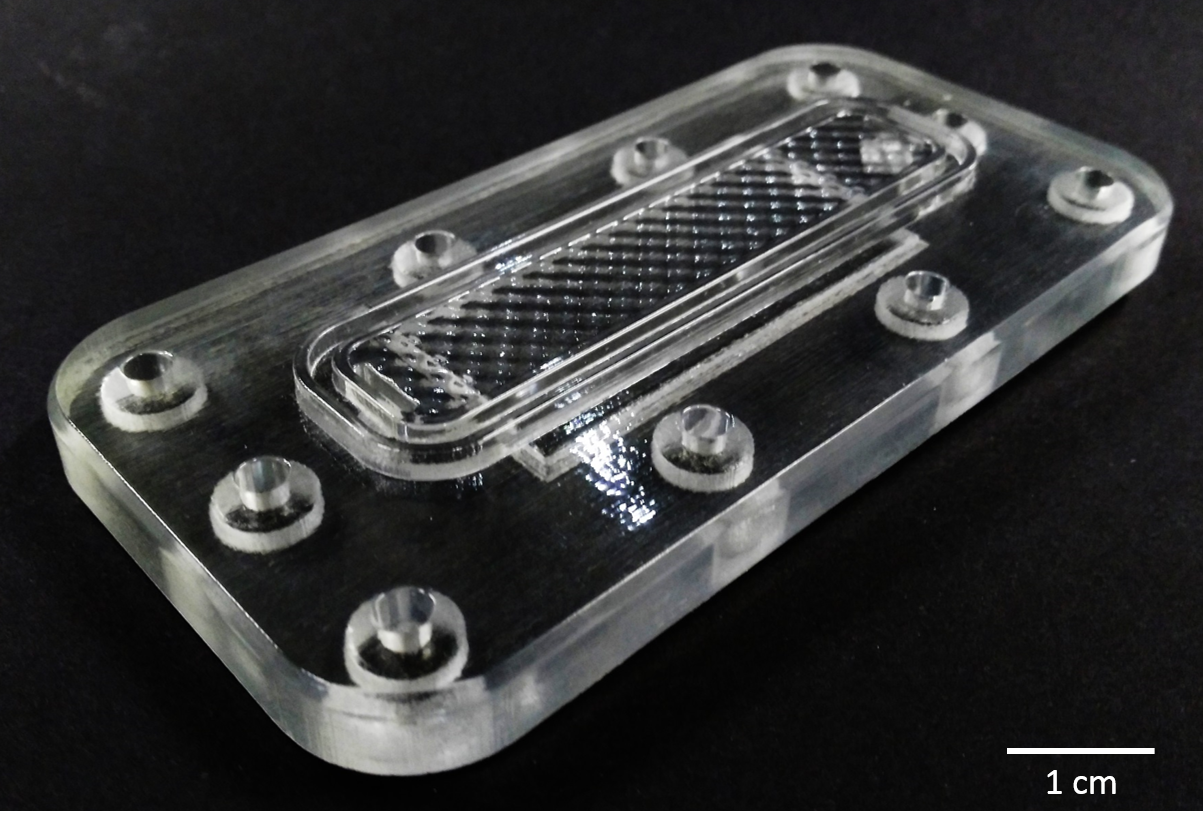}
    \caption{3D-printed membrane module with built-in sinusoidal corrugation.}
    \label{fig:3D_sinus}
\end{figure}

The permeation rate for different built-in spacer geometries was experimentally evaluated in an additively manufactured membrane module that was fabricated using polyjet 3D printing (Stratasys, Objet Eden 260V). The membrane module shown in Figure~\ref{fig:3D_sinus} was printed layer by layer with a transparent photopolymer (Stratasys, RGD810). During printing, the internal fluid channels need supporting structures (Stratasys, SUP705), which were later removed using a high-pressure washer (Krumm-tec, RK Top 5). Any remaining material was subsequently dissolved in a stirred bath of \SI{1}{\mole\per\liter} sodium hydroxide for approximately 24 hours.

\subsection{Design of the built-in spacer}\label{design}
For the conventional implementation, a mesh spacer, usually a fabric, would be placed in the empty feed channel, which also functions as an additional element to protect the membrane from damage. The advantage of replacing mesh spacers with built-in spacers is the possibility of single-step production that requires less time, e.g., injection molding, without the need for a spacer placement step. During injection molding, polymer granulate is molten and transferred into a metal shape injected into a mold cavity via a nozzle. For the design of a spacer built-in in the membrane module, any ramifications or undercuts need to be avoided in order to facilitate removing the molded part. Simple superficial cavities ensure a uniform distribution of the mold mass in the tool and a smooth surface of the module resulting in consistent product quality~\cite{Gibson2015}. Built-in spacer structures were designed following injection molding restrictions, thus avoiding ramifications or undercuts and highly detailed structures. 

\begin{figure}[H]
	\begin{center}
	  \includegraphics[width=\textwidth]{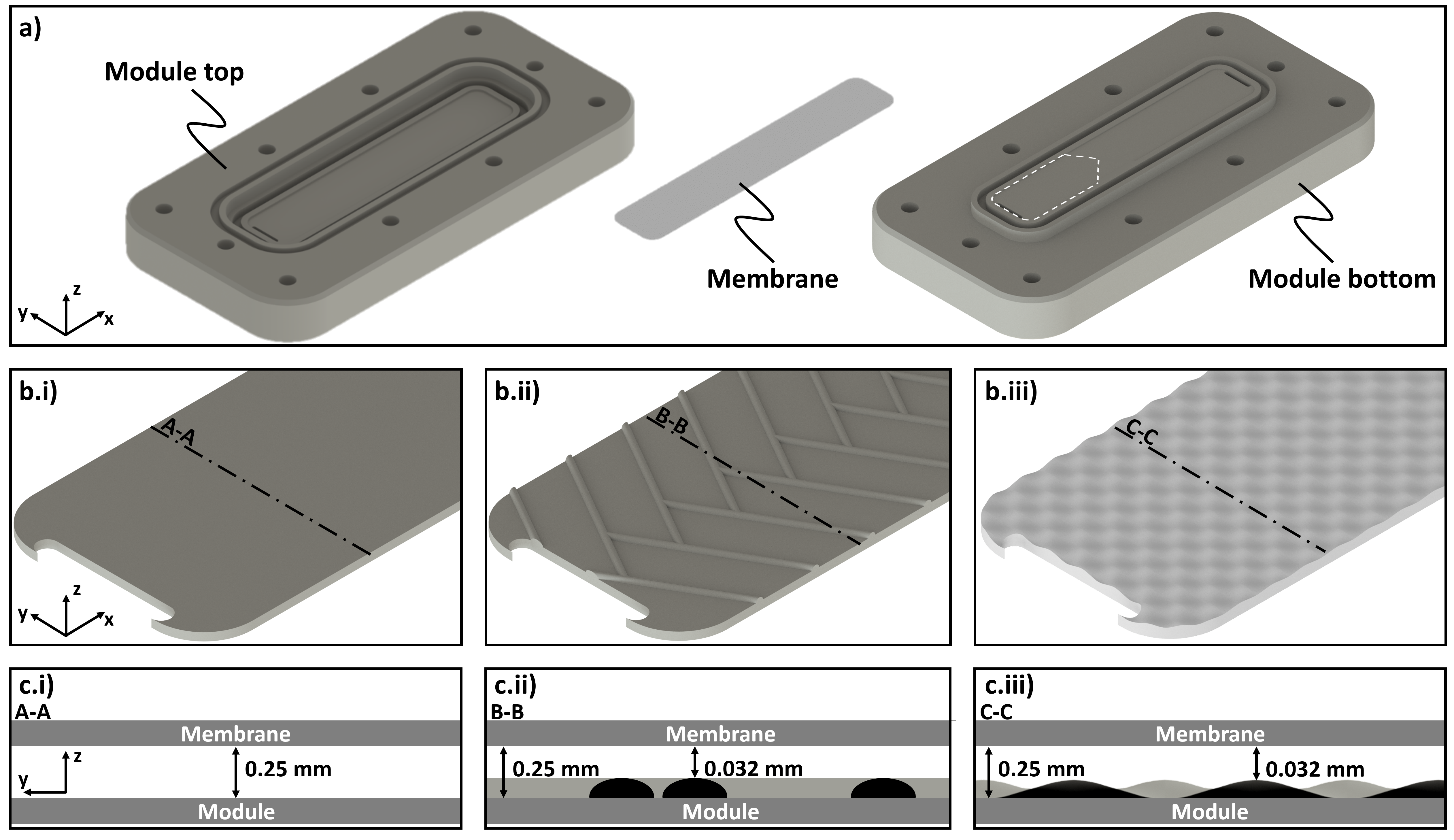} \caption{a) Module top with the permeate outlet, module bottom with the feed channel containing inlet and outlet openings and membrane to be placed in between. b) Feed structure for the b.i) empty channel, b.ii) the herringbone structure and the b.iii) Sinus{oidal corrugation}. c) Cross section of the three feed channels with dimensions.}
	  \label{fig:Uebersicht}
	\end{center}
\end{figure}

The module investigated in this work is a small-scale crossflow module consisting of two parts. Figure~\ref{fig:Uebersicht} shows the two module parts with the membrane in between (Figure~\ref{fig:Uebersicht}~a), the feed channel with and without the two final built-in structures (Figure~\ref{fig:Uebersicht}~b), and the corresponding geometrical dimensions (Figure~\ref{fig:Uebersicht}~c). The module top contains a permeate outlet but is not implemented in this work. The module bottom contains the feed inlet and a retentate outlet. The feed inlet leads to a rectangular channel upon which the membrane is placed. Inlet and outlets measure a diameter of \SI{1}{\milli\meter} each. The feed channel measures \SI{13,5}{\milli\meter} in width and \SI{76}{\milli\meter} in length. The empty feed channel has a height of \SI{250}{\micro\meter}, which is filled with different built-in structures. The total module measures around \SI{10}{\centi\meter} x \SI{3}{\centi\meter} in size. 

To find a suitable novel spacer design, the literature was studied, and new approaches were implemented. In the literature, results from 2D and 3D simulations, as well as from experiments, were compared. Based on these studies, ten structures were designed and investigated in laminar-flow CFD simulations, from which two final structures were selected (for extensive information on the discarded structures, see Supplement Figures~3 to 5). As geometric length to be kept constant for comparing the structures, the membrane distance was chosen here instead of the remaining void volume. The selection process was based on balancing the shear rate factors at the membrane, shear rate distribution, and pressure drop between inlet and outlet. The selected structures, a staggered herringbone and a sinusoidal corrugation, were compared to the empty channel in simulations and additionally to an inlay net spacer in experiments. 

\paragraph{Empty channel} 
Channel heights on the feed and permeate sides are chosen in accordance with the used mesh spacers. The investigated empty feed side channel was set to a height of \SI{250}{\micro\meter} as adapted to the lower limit for the height of so-called tight commercial biotechnological membrane modules.

\paragraph{Herringbones} Staggered herringbone structures are known to improve mixing in microfluidics normal to the flow direction inducing secondary flow patterns \cite{Wiese.2018}. They are studied in this work for different element distances and the element height. To form herringbones, rods of \SI{0.22}{\milli\meter} radius and \SI{32}{\micro\meter} membrane distance are arranged at an angle of \SI{60}{^{\circ}} and \SI{3}{\milli\meter} of distance. The dimensions result in 22 rods along the feed channel.

\paragraph{Sinusoidal corrugation} As geometries to induce secondary flow patterns, sinusoidal surface structures as part of membrane modules are studied to a much lesser extent as they cannot be created by filament-based spacers. To create the sinusoidal surface, a basic element with four perpendicular sides is constructed (amplitude $A=\SI{0.109}{\milli\meter}$, wave length $\lambda=\SI{2}{\milli\meter}$). The adjacent sides of the basic element display sinus functions, which differ by half a period. As the sinus functions are orthogonal, they form a cross-shaped pattern. A complete sinus wave measures \SI{2}{\milli\meter} and, like the herringbone structure, has a distance to the membrane of \SI{32}{\micro\meter}. These dimensions result in 38 peaks for the feed channel length. As a variation of the sinusoidal corrugation, the number of peaks is halved. 

\subsection{Fluid dynamics simulation} \label{MM_CDF}
3D-CFD simulations using COMSOL Multiphysics were performed during the design process of the membrane module to create an optimized spacer geometry. In particular, the flow conditions at the membrane-fluid interface were optimized with the geometry of the built-in spacers. Flow simulations served for a pre-selection aiming for a homogeneous flow field and were conducted for ten spacer geometries (see Supplement Figures~3 to 5). Different spacer configurations were simulated with laminar inflow conditions and ambient pressure at the outlet. The chosen material for the fluid properties was pure water from the COMSOL material library.

This study focuses on the holistic process of designing an built-in spacer structure suitable for injection molding. Thus, the model was run at a stationary operating point, and the inlet velocity was set to \SI{0.152}{\meter\per\second}, which equals a flow of \SI{6.7}{\milli\liter\per\minute} passing the inlet cross-section of \SI{1}{\milli\meter} diameter. This flow condition results in a Reynolds number of 18 for the empty feed channel. Since a stationary flow field was expected, steady-state conditions were chosen for the solver.

The model used a constant temperature of \SI{293.15}{\kelvin}. All simulations assumed a fully developed and constant feed flow. Calculations were done using the internal PARDISO solver, obtaining fully coupled conditions, and direct iterations. 

Geometrical adjustments (e.g., symmetry) and periodic boundary conditions were applied to the simulations to reduce the calculation effort. For the structures presented here, only the empty channel was implemented with symmetry. To accurately resolve the boundary layer at the membrane surface, a structured mesh was used, combined with unstructured tetrahedral elements in the remaining region, resulting in a total of $1.1E5$ mesh elements (empty channel). The mesh was refined based on a stable pressure loss probed between the inlet and outlet. The implemented boundary conditions and a plot of the mesh independence study are included in Supplement Figures~1 and 2.

We used diluted-species retention to investigate and represent the effect of flow conditions on solute retention for the different geometries. This work implemented \textit{transport of diluted species} to describe mass transport and thus concentration polarization at the membrane-fluid interface. To model concentration polarization and guide the fabrication process, continuous-flow conditions, which are the case for reverse osmosis systems, were introduced. Ultrafiltration and reverse osmosis can be considered similar regarding mechanisms creating concentration polarization, as both are dominated by advection of the liquid medium~\cite{Matthiasson1980}. Thiess et al. also used the similarity between ultrafiltration and reverse osmosis phenomena~\cite{Thiess2017}. This study used specifications for a reverse osmosis membrane and its permeability for the chlorophenol concentration studied by Al-Obaidi and Mujtaba~\cite{AlObaidi2016}, called RO-approach in the following. The calculated results of Al-Obaidi and Mujtaba were validated by experiments~\cite{Sundaramoorthy2011}.

Combining laminar flow with transport of diluted species prolongs the simulation time significantly as it combines convective flow phenomena with diffusive transport. For this study, the following assumptions were made:
\begin{itemize}
	\item membrane rejection of 100\,\% for solutes;
	\item membrane modeled as rigid element without deformation due to pressure;
	\item constant ambient pressure and pure water on permeate side;
	\item constant feed composition with a fixed solute concentration;
	\item and density, viscosity, and diffusion coefficients are constant and concentration-independent.
\end{itemize}

Transport of diluted species is dominated by convective flow according to the Navier Stokes equations. Concentration gradients cause species transport in the form of diffusion. In addition to flow tangential to the laminar streamlines, diffusive mass transport can also occur normal to the flow direction. In this study, mass was neither produced nor consumed nor changed over time. Thus, the total fluid movement was dominated by convective and diffusive flow.

To enable permeate volume flow, trans-membrane pressures (TMP) were applied by setting $p_{Outlet}$ as outlet condition in the laminar flow node and to later calculate the TMP according to: 
\begin{equation}
\text{TMP} = \frac{p_{\text{Outlet}} + p_{\text{Inlet}}}{2} - p_{\text{Permeate}}.
\label{eqn:TMP}
\end{equation}
The outlet pressure on retentate side was set between $\SI{0}{\bar}$ and $\SI{12.5}{\bar}$ in steps of $\SI{2.5}{\bar}$. The model evaluated the corresponding inlet feed pressure, which resulted to be slightly higher than the outlet pressure. The pressure on permeate side was at ambient pressure, which equals an overpressure of zero in this model. Thus, the local pressure p corresponds to the pressure difference between the feed and permeate side, acting as driving force. 

The addition of a diluted species demanded a further characterization of the mass transport at the membrane. The model provided a permeate flux as a second outlet condition as function of the pressure difference and the concentration at the membrane. It was described as the orthogonal outlet velocity $U_{p}$, where the characteristics of the membrane are summarized in the permeability or solvent transport coefficient $A_{w}$: 
\begin{equation}
	U_{p} = A_w \cdot \left(p-\left(R \cdot T \cdot c\right)\right).
\label{eqn:u_perm}
\end{equation}
The membrane flux increases with higher pressure difference $p$ and decreases with increasing osmotic pressure difference $\pi = R \cdot T \cdot c$ due to higher local concentration (see equation (\ref{eqn:u_perm})). 
To compare the different structures, the permeate volume flow was determined by integrating the orthogonal outlet velocity over the membrane surface for the different TMP. A sketch of the geometry with the applied boundary conditions can be found in the supporting information (see Supplement Figure~1).

Flow conditions and concentration-related assumptions taken from \cite{Sundaramoorthy2011} are summarized in Table~\ref{tbl:RO_paper}, where $c_0$ stands for the entry concentration, $c_p$ for the permeate concentration, and $A_w$ for the membrane permeability.
	\begin{table}[h!bt]
		\begin{center} 
			\caption[System conditions for simulation with concentration polarization]{System conditions for simulation with concentration polarization using material properties according to \cite{Sundaramoorthy2011}.} \vspace{2ex} 
			\begin{tabular}{lll}         
			\hline  
			\multicolumn{1}{l}{Parameter} & \multicolumn{1}{l}{Value}&\multicolumn{1}{l}{Unit}\\
			\hline
			\multicolumn{3}{l}{\textbf{Laminar flow}}\\
			$v_{in}$ & \SI{0.152}{}&\SI{}{\meter\per\second}\\
			$p_{outlet}$ & 0; 2.5; 5; 7.5; 10; 12.5&\SI{}{\bar}\\ \\
			\multicolumn{3}{l}{\textbf{Transport of diluted species}}\\
			$c_0$ & \SI{6.23}{}&\SI{}{\mole\per\cubic\meter}\\
			$c_p$ & \SI{0}{}&\SI{}{\mole\per\cubic\meter}\\
			$D$ & \SI{1.5E-9}{}&\SI{}{\square\meter\per\second}\\
			\\
			\multicolumn{3}{l}{\textbf{General}}\\
			$A_w$ & \SI{9.3943E-12}{}&\SI{}{\square\meter\second\per\kilo\gram}\\
			Diluted component & Chlorophenol &\\
			\hline
			\end{tabular}
			\label{tbl:RO_paper}	
		\end{center}
\end{table}

\subsection{Experimental characterization of permeate flux}
Simulation results determined the most promising spacer structures regarding shear rate/ pressure drop ratios and motivated those to be 3D-printed and tested in experiments with protein filtration. The proposed built-in spacer structures were supplemented by and compared with commercial net spacers. \\

The tested ultrafiltration membrane was a Hydrosart\textsuperscript{\textregistered} 30 kD by Sartorius Stedim GmbH. It was placed either in an empty 3D-printed prototype or in 3D-printed modules with built-in structures. In the empty module, a  Clearedge \SI{230}{\micro\meter} acted as a benchmark mesh-type spacer, which was placed on top of the membrane. Six screws then fixed the module housing at a torque of \SI{1}{\newton\meter} with a torque wrench (DremoMeter DBGM Rahsol 10-100 cm kp) to ensure proper sealing.

Due to the preservation of the membrane with glycerin, each module was first flushed with \SI{100}{\milli\liter} ultrapure water. The inlet feed volume flow was set to \SI{10}{\milli\liter\per\minute for flushing}. All modules were tested for integrity and water flux beforehand. The water flux test included flushing of \SI{100}{\milli\liter} solution while tracking the permeate. Only modules with a specific, membrane-related permeate quantity according to the datasheet were used for the experiments.
Water flux test and experiments were performed at a trans-membrane pressure of \SI{1.5}{\bar} while the permeate channel remained open to ambient pressure (retentate outlet was closed for static water flux tests). The module operated vertically from bottom to top to prevent gas bubble entrapment during filling. During the experiments, the retentate and feed pressure, the volume flow, and the permeate amount were recorded. Permeate volume flow was measured simply by using a stopwatch and a measuring cylinder. As a model protein, BSA (CAS 9048-46-8, Carl Roth) was used to prepare a \SI{10}{\gram\per\liter} and a \SI{50}{\gram\per\liter} solution in \SI{10}{\milli\mole} phosphate buffer of pH 7, pre-filtered with \SI{0.2}{\micro\meter} filter. The experimental investigation was designed to emulate a later application of approx. 6-8h process run time. 

\subsection{Permeation setup and (flow-) MRI measurements}
The technique of magnetic resonance imaging (MRI) allows the visualization of structures and fluid flow inside membrane modules in-situ and non-invasively. It utilizes the property of nuclear spins in water aligned to a magnetic field and decaying in time. Using appropriate sequencing of magnetization and signal read-out allows for analyzing the displacement of fluid elements (voxels). In this study, the measurements were performed on a Magritek low-field (LF) NMR tomography system and a Bruker BioSpec 70/20 USR high-field (HF) Scanner. While the LF system operates at a field strength of \SI{0.56}{\tesla}, the HF system operates at \SI{7}{\tesla}. A 2D spin-echo pulse sequence was used in all measurements. MRI parameters are provided in Table~\ref{tab:MRISettings}. The signal-to-noise ratios in all measurements were larger than~8.  Post-processing was performed with \textsc{Matlab}. The measuring technique is explained in more detail in our previous studies~\cite{Wypysek2019,Luelf.2018,Wiese.2018b, Wiese.2019, WYPYSEK.2020}.

Permeation experiments were conducted in a commercially available \SI{30}{\kilo Da} ambr\textsuperscript{\textregistered} crossflow module modified as a screening filter for variable concentrations up to viscosities above \SI{50}{\centi p} and a broad range of molecules. A constant feed flow of \SI{10}{\milli\liter\per\minute} in the LF experiments and \SI{5}{\milli\liter\per\minute} in the HF experiments of a buffer solution consisting of de-ionized water and \SI{1.48}{\gram\per\liter} copper(II) sulfate pentahydrate (\ce{CuSO4.5H2O}, CAS:~7758-99-8, Carl Roth, Germany) was set using two continuous neMESYS syringe pumps by CETONI GmbH, Germany. A bubble trap was used to prevent air from entering the module. The pressure in the retentate channel was set to \SI{1.5}{\bar} over-pressure using a back-pressure valve. 

\begin{table}[H]
    \centering
    \caption{MRI parameter settings for the flow measurement of the ambr\textsuperscript{\textregistered} crossflow module for the LF and HF tomograph.}
    \begin{tabular}{llll}\hline
 & Value LF & Value HF  & Unit \\
 &  &No flow / flow  &  \\\hline
    Repetition time              & 400   & 1000 / 200 & \SI{}{\milli\second}                          \\
    Echo time                    & 32   & 29 / 5 & \SI{}{\milli\second}                          \\
    Spatial res. [x $\cdot$ y]& 30 $\cdot$ 8 & 25 $\cdot$ 5  &  \SI{}{\cubic\milli\meter}\\
    No. of pixels [x $\cdot$ y]  & 256 $\cdot$ 128 & 400 $\cdot$ 80 &  \SI{}{-} \\
    Slice thickness & 10 & 0.5 &\SI{}{\milli\meter} \\
    Number of averages              & 12 & 5 / 56  &  \SI{}{-}  \\ \hline
    \end{tabular}
    \label{tab:MRISettings}
\end{table}

\section{Results and discussion}
\subsection{Laminar flow simulations visualize hydrodynamics}
In addition to vortex formation, shear stress is also considered to favor mass transport \cite{Schwinge2002b}. Shear stress is caused by a tangential force generated through distortion of fluid and can reduce particle deposition on membrane and module surfaces. Thus, high shear rates are considered to increase the permeation rate of an ultrafiltration membrane \cite{Cao2001, Lee2013}.
 
\begin{figure}[H]	
	\begin{center}
	  \includegraphics[width=\textwidth]{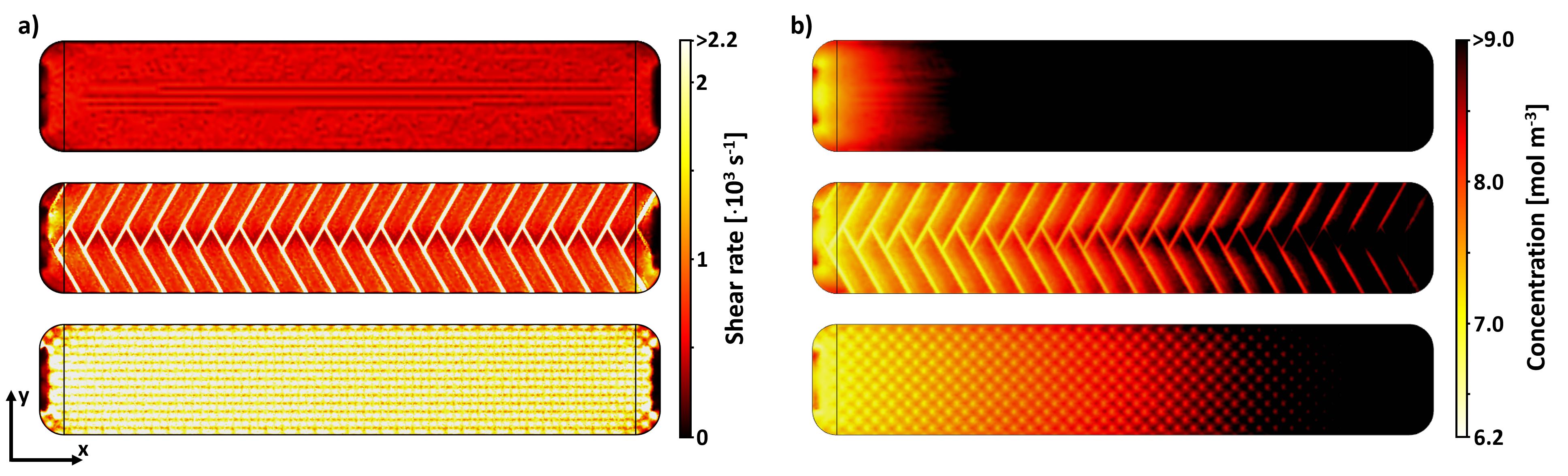} \caption{a) CFD simulations show (top) even but low shear rate distribution for the empty channel, (middle) shear rate peaks above the herringbone structures, and (bottom) various peaks and valleys for the sinusoidal corrugation. b) (top) In the empty channel, the concentration increases fast due to concentration polarization. (middle) The herringbone structure delays the concentration increase. (bottom) The Sinus proposes the slowest concentration increase and an even distribution.}
	  \label{fig:sim_ConcSr}
	\end{center}
\end{figure}

The study below compares spacer designs in order to identify high shear rate distributions averaged over the membrane. Figure~\ref{fig:sim_ConcSr} shows the development of shear rate and concentration over the whole module (inlet on the left; outlet on the right) for the empty channel, herringbone structure and sinusoidal corrugation. In both Figure~\ref{fig:sim_ConcSr}~a) and b), darker colors indicate the undesired case (low shear rate; high concentration). The CFD results show that the highest shear rates (shown in white in Figure~\ref{fig:sim_ConcSr}~a)) are achieved for small distances between membrane and module wall. The empty channel (Figure~\ref{fig:sim_ConcSr}~a) (top)) shows an even shear rate distribution with low shear rates around \SI{0.5E3}{\per\second}. For spacer designs, where the shape runs over the whole module width, shear rate peaks appear above the top area of the spacer shape. This is the case for the herringbone structure (Figure~\ref{fig:sim_ConcSr}~a) (middle)), where extremely high shear rates up to \SI{3E4}{\per\second }are observed over the diagonal bars with \SI{32}{\micro\meter} membrane distance compared to low shear rates not higher than \SI{1.5E3}{\per\second} (in red) in between those bars (\SI{250}{\micro\meter} membrane distance). 

In the case of the sinusoidal corrugation, the filaments do not form a continuous barrier over the whole width; instead, alternating peaks and valleys appear. The varying structure results in an even distribution and high shear rates (around \SI{2E3}{\per\second}, see Figure~\ref{fig:sim_ConcSr}~a) (bottom)), simultaneously. Within the literature, the shear rate values are well comparable with CFD simulations of mesh-type spacers and similar inlet velocities~\cite{Karode2001}. In general, the trade-off for the increased wall shear stress is an increase in pressure drop along the feed channel \cite{Schwinge2002, DaCosta1991,DaCosta1993, DaCosta1994}. However, in this case, the sinusoidal corrugation does not elevate the pressure loss (see Supplement Figure~5).

In the literature, wall shear stress is often considered. In this study (Re~=~18, TMP $\SI{12.5}{\bar}$), a value of $\SI{1.7}{\pascal}$ is obtained, which correlates with simulated values of Mazinani et al. \cite{Mazinani.2019} and is in the same order of magnitude compared to recommendations for optimal operation of common UF applications (2~-~6~Pa, see \cite{Chew2004}).

The shear rate highly depends on the local velocity. The spacer structures occupy different amounts of space in the empty channel, consequently narrowing the cross-section. This leads to a higher flow velocity and thus to an increase in shear rate. The void volume of the empty channel measures \SI{332}{\cubic\milli\meter}, which is a lot larger than that of the herringbone structure (\SI{306}{\cubic\milli\meter}) and the sinusoidal corrugation (\SI{223}{\cubic\milli\meter}) and leads to the observed shear rate increase. The inserted structures were not normalized regarding the remaining void space but instead kept a fixed membrane distance in order to maintain similar conditions for the shear rate effects. The proposed membrane distance of \SI{32}{\micro\meter} resulted from previous investigations.

\subsection{Diluted species retention simulations}
\label{sec:SoluteSimulation}

To investigate the correlation between shear rate and concentration, 2D plots for the concentration on the membrane surface were compared (see Figure~\ref{fig:sim_ConcSr}~b)). Low concentrations at the membrane are desired as they result in a small osmotic pressure, which, in turn, increases the permeate flux (see the function for the permeate flux in equation (\ref{eqn:u_perm})). In the legend in Figure~\ref{fig:sim_ConcSr}~b), light colors indicate low concentrations. 
Figure~\ref{fig:sim_ConcSr}~b) shows a concentration increase from left to right, which correlates with the feed-flow direction. The concentration trend implies an increase in the amount of the retained component. 
The empty channel (Figure~\ref{fig:sim_ConcSr}~b) (top)) shows rapid growth in concentration at an early position within the first eighth. The herringbone and the sinusoidal corrugations (Figure~\ref{fig:sim_ConcSr}~b) (middle) and (bottom)) show that high concentrations are reached at a later point: the herringbone structure reaches the same concentration level as the empty channel at halfway, whereas the sinusoidal corrugation delays the concentration increase to the last third. 
As in the case of shear rates, the lowest concentrations (indicated by light colors) are observed on top of obstacles: above bars for the herringbone structure and above sinus peaks for the sinusoidal corrugation. This observation correlates well with other studies~\cite{Santos2007}. Where there were high shear rates above the filaments of the herringbone structure in Figure~\ref{fig:sim_ConcSr}~a) (middle), there are now low concentrations in the same regions (lighter colors compared to the space between filaments). The sinusoidal corrugation represents a more even concentration distribution compared to the herringbone structure.\\

\begin{figure}[H]
\begin{centering}
\includegraphics[width=0.85\textwidth]{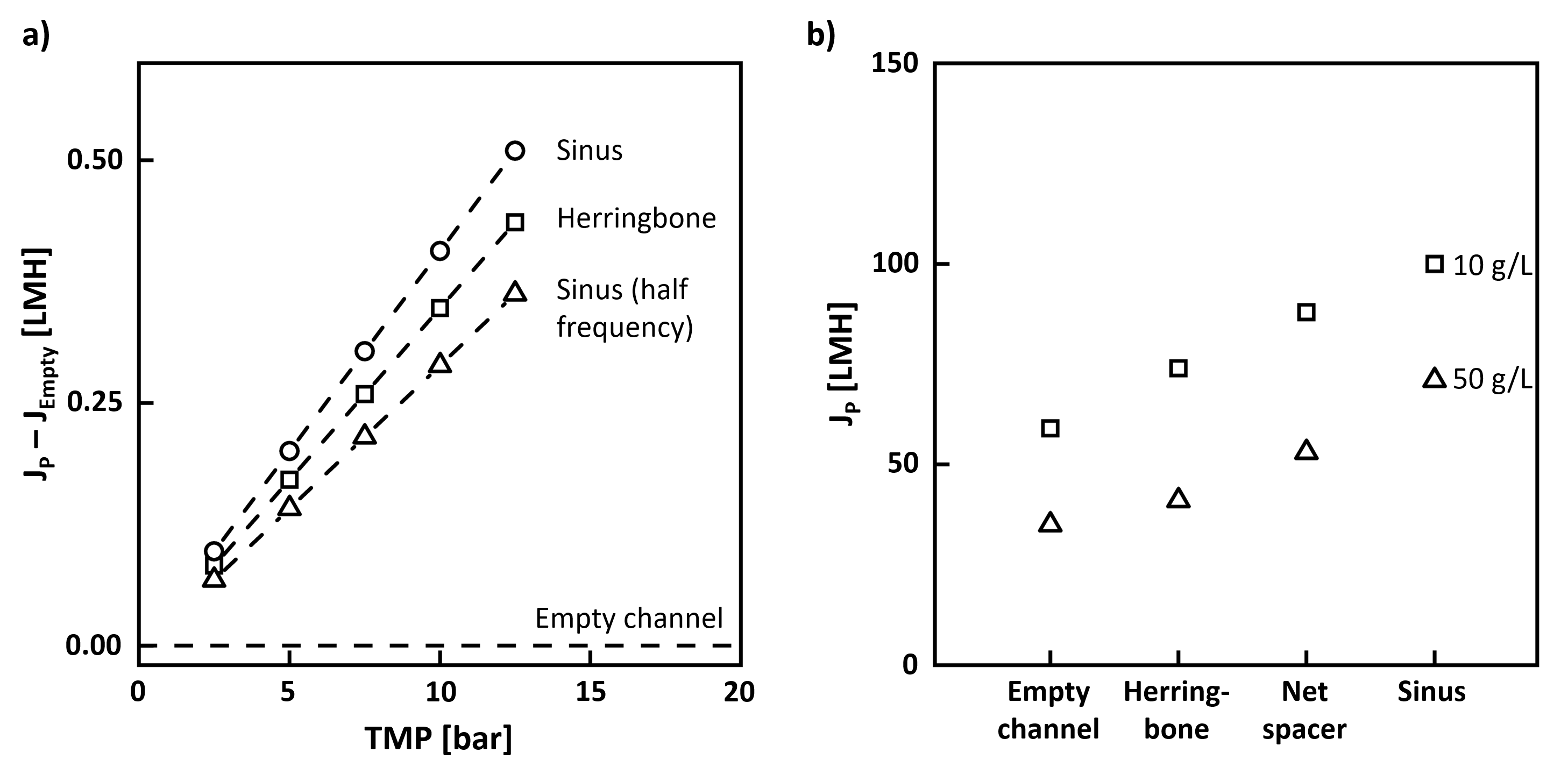}
\caption[]{a) Simulations show the highest permeate flux enhancement for the sinusoidal corrugation, followed by herringbone structure and the sinusoidal corrugation with half the frequency. All results are better than those for the empty channel. b) sinusoidal corrugation achieves the highest permeate flux in BSA experiments, followed by the commercial net spacer, herringbone structure, and the empty channel.}\label{fig:BSAandFlux}
\end{centering}
\end{figure}

Figure~\ref{fig:BSAandFlux}~a) shows the simulated permeate fluxes of the sinusoidal corrugation and the herringbone structure in comparison to the empty channel. A second sinusoidal corrugation with half the frequency, and thus half the number of sinus peaks, demonstrates lower permeate flux differences than the sinusoidal corrugation but outperforms the herringbone structure. The simulation model specifies an inlet velocity and different outlet pressures since these boundary conditions are relevant in experiments (see section~\ref{MM_CDF} and Supplement Figure~1). Thus, permeate fluxes, velocity and pressure profiles develop according to the respective structure. The permeate flux rises for increasing TMP according to equation (\ref{eqn:u_perm}). As this study uses RO as a replacement system, the permeate flux is small compared to an ultrafiltration process, which complicates visualizing differences in the respective permeate fluxes clearly. Thus, the TMP is increased up to $\SI{12.5}{\bar}$, which is also in accordance with pressures used by Al Obaidi et al. \cite{AlObaidi2016}. The permeate fluxes of the spacer structures $J_P$ are normalized to TMP and corrected by the permeate flux of the empty spacer channel $J_{empty}$ for different TMP according to $J_{compare}=J_P-J_{empty}$ (see Figure~\ref{fig:BSAandFlux}~a)). A simulated permeate flux of $\SI{3.2}{LMH\per\bar}$ obtained for the sinusoidal corrugation is in the same order of magnitude compared to Al Obaidi et al. (roughly calculated to $\SI{0.26}{LMH\per\bar}$ with given parameters and graphs \cite{AlObaidi2016}). 

\begin{figure}[H]
    \begin{center}
    \includegraphics[width=0.90\textwidth]{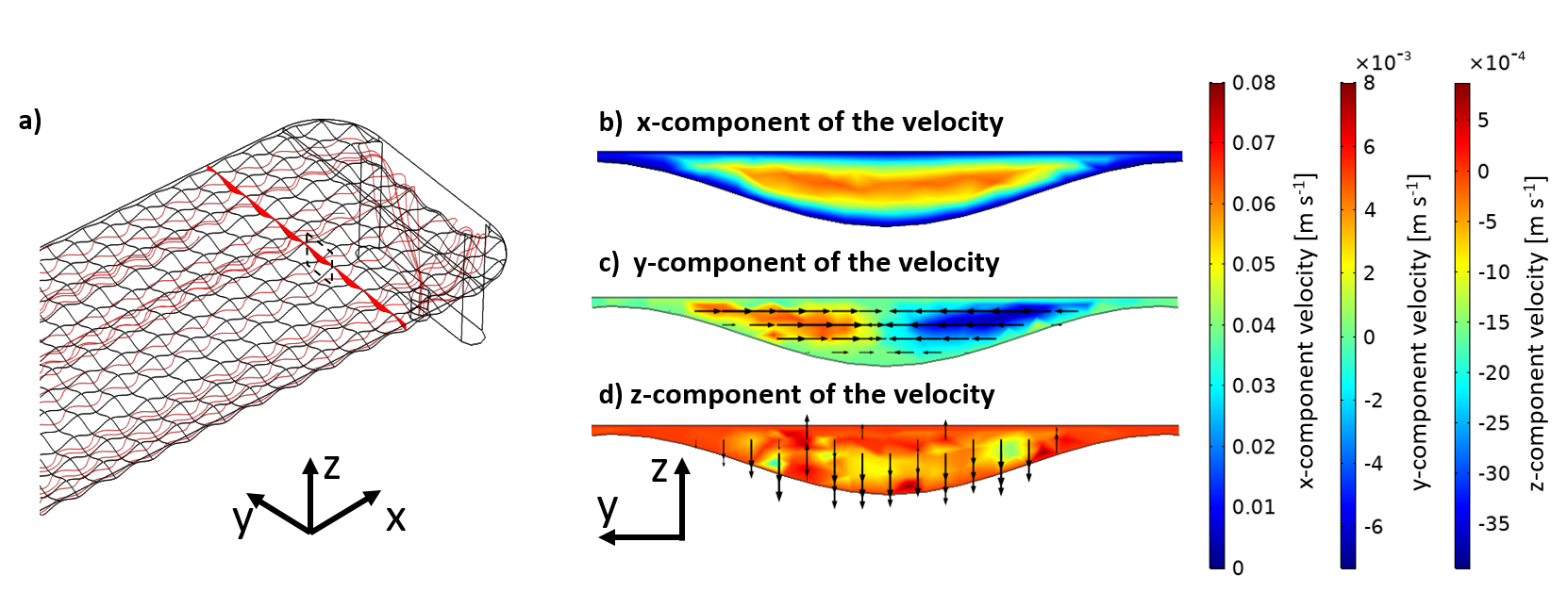}
\caption[]{a) View of the cut plane in flow direction and streamlines showing the flow around the sinusoidal corrugation. b-d) The x-component of the velocity shows the main fluid flow, whereas the y- and z-component reveal a change in flow direction induced by the sinusoidal corrugation.}\label{fig:velocityprofile}
\end{center}
\end{figure}

Figure~\ref{fig:velocityprofile}~a) shows the feed outlet of the module with streamlines indicating the flow being redirected by the sinusoidal corrugations. A close-up of a sinus bulge of the indicated cut plane in y-z (see the mark in the red plane in Figure~\ref{fig:velocityprofile}~a)) illustrates the x-, y- and z-component of the velocity (see Figure~\ref{fig:velocityprofile}~b)-d)). Figure~\ref{fig:velocityprofile}~b) shows higher velocities than Figure~\ref{fig:velocityprofile}~c) and d), indicating the main flow in the x-direction. The velocity components and arrow plots in Figure~\ref{fig:velocityprofile}~c) show that the fluid moves around the sinusoidal peak resulting in positive and negative movement in the y-direction. Alternating sinusoidal corrugations up- and downstream induce reversing flow in z-direction indicated by a change of sign in Figure~\ref{fig:velocityprofile}~d). The illustrations in Figure~\ref{fig:velocityprofile}, streamlines along the x-axis (see Supplement Figure~6), and the low Reynolds number of 18 imply a laminar flow profile rather than vortex formation. Thus, the higher permeation rate of the sinusoidal corrugation and better mixing can be justified by the lateral movement shown in Figure~\ref{fig:velocityprofile}~c) and d), narrowing streaming layers of different concentrations and thus favoring diffusion. In their review on 3D simulations with spacers, Fimbres-Weihs, and Wiley \cite{FimbresWeihs2010} commented that 3D structures induce movement perpendicular to the flow direction and thus enhance mass transfer, which also applies to the study at hand. Mazinani et al.~\cite{Mazinani.2019} also implemented a sinusoidal structure, but on the membrane side. Their structure is similar to our sinusoidal corrugation ($A=\SI{0.5}{\milli\meter}$, $\lambda=\SI{3}{\milli\meter}$~\cite{Mazinani.2019}; for sinusoidal corrugation see section~\ref{design}). Although Mazinani et al. have visualized vortex formation in between sinusoidal peaks in 2D simulations, the crucial difference to our case is the distance between membrane and module housing of $\SI{4}{\milli\meter}$ height, which allows for extensive flow formation. In our case, a maximum feed channel height of $\SI{0.25}{\milli\meter}$ rather tends to redirect the laminar flow shallowly.

The critical Reynolds number, usually an indicator for the transition from steady to unsteady flow, can be identified for complex, realistic spacers~\cite{Qamar2019} or calculated for rather simple structures~\cite{Santos2007, Schwinge2002}, and lies between 100 and 400, which exceeds the here obtained Reynolds number of 18. Further simulations with Reynolds numbers up to 2500 were carried out but did not show any transition to unsteady flow (see Supplement Figure~6). The simulations revealed a relation to the power of 1.73 between pressure drop and velocity, which is in good agreement with other exponents modeled in the literature for the pressure drop-velocity relationship typical for spacers (1.63~-~1.99 in~\cite{Karode2001}, 1.69~-~1.82 in~\cite{DaCosta1991}, 1.66~-~1.84 in~\cite{DaCosta1994}). An exponent of 1.73 indicates turbulent flow, but mass transport can also develop laminarly despite the reported exponents in mesh-type spacers~\cite{DaCosta1991}.

High pressure drop and high volume flow both affect pumping performance in the later application, especially when scaling up to larger membrane modules. A substantial increase in pressure drop is undesirable because it leads to higher operating costs, e.g., due to pumping capacity~\cite{DaCosta1991}. Given that the module is implemented as a single-use product for ultrafiltration processes, peripherals (vessels, hoses) are also designed to be disposable and often limited to maximum operating pressures of 3~-~4~bar. 
Although higher volume flow usually has a positive effect, turbulence can also negatively affect friction losses and heat input and lead to higher operating costs (especially regarding the pumping capacity at high volume flows and high viscosities, as in UF processes). Turbulence is especially undesired in the main flow due to friction losses and temperature increase, in addition to the almost quadratic pressure drop mentioned above. The temperature increase is equivalent to energy dissipation and can negatively affect the products present (proteins, enzymes). For this reason, a low-pressure drop was aimed at in the pre-simulations (see Supplement Figure~5).

\subsection{Experimental validation from BSA experiments} \label{Res_Exp}
Figure~\ref{fig:BSAandFlux}~b) shows the permeate volume flow $J_P$ for the empty channel, the herringbone structure, the sinusoidal corrugation (all 3D-printed prototypes), and a commercial net spacer, which was placed in the empty membrane module. Permeate volume flows were measured for BSA concentrations of \SI{10}{\gram\per\liter} and \SI{50}{\gram\per\liter}. Figure~\ref{fig:BSAandFlux}~b) shows that the permeate volume flow is higher for low BSA concentrations, according to equation~(\ref{eqn:u_perm}). The sinusoidal corrugation shows the best performance with the highest permeate flow rate, followed by the commercial net spacer, the herringbone structure, and the empty channel. In Figure~\ref{fig:BSAandFlux}~b), the permeate flux increase of the sinusoidal corrugation is approximately \SI{10}{\percent} to \SI{30}{\percent} compared to the commercial net spacers, reaching absolute values of $\SI{40}{LMH\per\bar}$ to $\SI{56.6}{LMH\per\bar}$. In comparison to the literature, these values are in the same range as in experiments under similar conditions conducted by Mazinani~et~al. (permeate flux $\SI{10}{LMH\per\bar}$ to $\SI{25}{LMH\per\bar}$ at Re~=~400, $\SI{1}{\gram\per\liter}$ BSA solution, see~\cite{Mazinani.2019}) and Thiess~et~al. (permeate flux $\SI{3.3}{LMH\per\bar}$ to $\SI{56.7}{LMH\per\bar}$, $\SI{100}{\gram\per\liter}$ BSA solution, see~\cite{Thiess2017}). Yet, in comparison with our simulations, the experimentally conducted permeate volume flow of the UF prototype module is around six times larger than the simulated amount based on the RO-approach (see section~\ref{MM_CDF}) used as a replacement system here. This can be explained by the fact that UF permeate fluxes exceed those of RO processes due to the more porous membrane and the working principle. This results in large permeate-flux differences between the two systems.

The BSA experiments demonstrate that all three spacer structures outperform the empty channel, which is in accordance with literature~\cite{Schwinge2002,Haidari2016}. Sinusoidal corrugations outperform the other spacer structures in agreement with the 3D-CFD simulations. Still, 3D-CFD simulations serve the qualitative comparison of the emerging trends rather than an exact match of the conducted experiments. In literature, mesh-type spacers are discussed to negatively influence biofouling due to their unfavorable shear stress distribution behind mesh-strand intersections~\cite{Koutsou2015}, and thus biofouling being rather an inevitable feed spacer problem~\cite{Vrouwenvelder2009}. Here, the built-in spacer structures are implemented as a superficial shape and therefore offer a positive effect compared to conventional mesh-type spacers. Although, in-situ biofilm formation has not been studied here, we assume that initial nucleation is unlikely to occur on the elevations of the sinusoidal corrugation but rather on the membrane. 

\subsection{Flow-MRI shows mixed fluid flow distribution}

\begin{figure}[H]
    \begin{center}
    \includegraphics[width=0.99\textwidth]{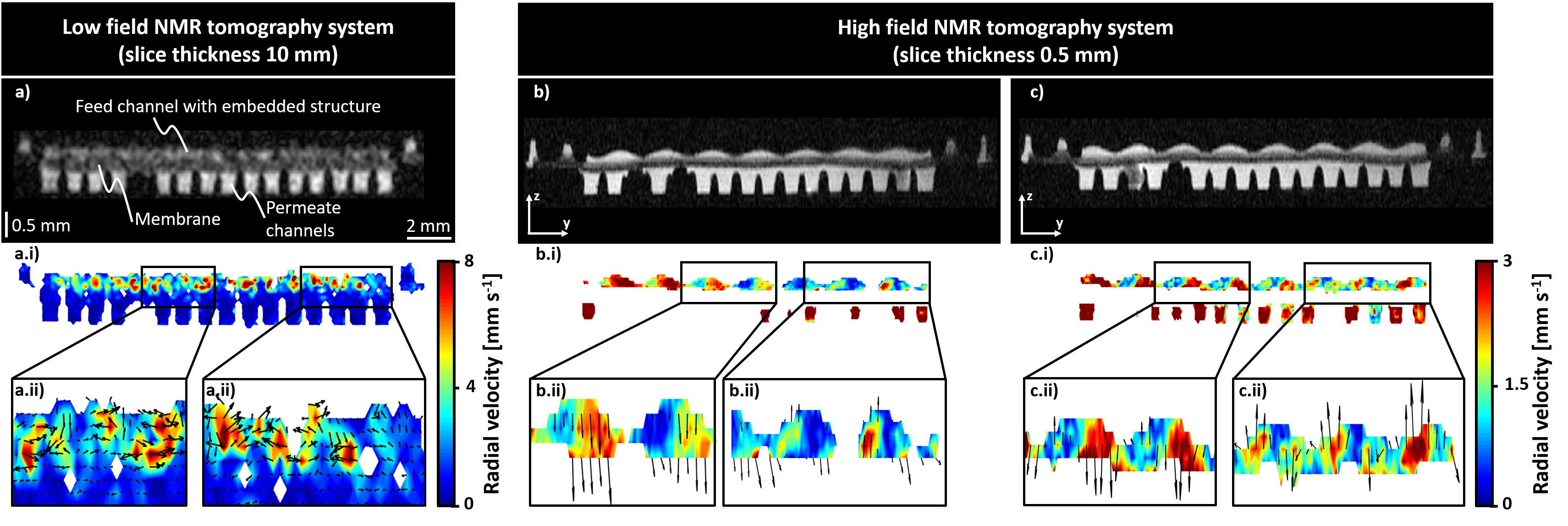}
\caption[]{(Flow-)Magnetic resonance imaging on an ambr\textsuperscript{\textregistered}
 crossflow module. (a) Low-field imaging, and (b) and (c) high-field imaging. Top row: MRI reveals the inner structure of the module. Middle row: Flow-MRI shows high radial velocity magnitude and vortex formation in the sinusoidal structured feed channel for the LF measurements, and straight and ordered flow in the HF measurements. Bottom row: Magnification shows flow directions inside the sinusoidal structure.}\label{fig:MRI}
\end{center}
\end{figure}

To better understand the permeate increase of the module with the sinusoidal corrugation, flow-MRI measurements under constant flow were conducted. Figures~\ref{fig:MRI}~a) depict cross-sectional images obtained by LF measurements. Figures~\ref{fig:MRI}~b) and c) show two different HF measurements of the module's cross-section with a distance of \SI{0.8}{\milli\meter} to each other. MR images (MRIs) in the top row in Figure~\ref{fig:MRI} reveal the inner structure of the membrane module, consisting of sixteen permeate channels, a membrane lying on top of the channels, and the feed channel with the built-in sinusoidal corrugation. The brighter the color in the magnetic resonance image, the higher the proton density in the respective pixel in the image, and thus, the water content is higher. The reason for permeate channels not being visible is the absence of water in the measuring area (see the fifth permeate channel in Figure~\ref{fig:MRI}~a) or the third and fifth permeate channel in Figure~\ref{fig:MRI}~b)). In the MRIs, the benefits of a high-field tomograph become obvious. LF imaging needs a bigger slice thickness to achieve high signal-to-noise ratios, which are desired. However, small structural details can no longer be distinguished, as the signal is averaged over the whole slice thickness. On the other hand, HF imaging can capture a higher level of detail with a smaller slice thickness.  

The middle row shows magnetic resonance flow measurements of radial velocities (on-plane velocities, see Figures~\ref{fig:MRI}~a.i) to c.i)) inside the membrane module with magnifications in the bottom row (see Figures~\ref{fig:MRI}~a.ii) to c.ii)). The color bar identifies the velocity magnitude, whereas the arrows represent the direction of flow. In the LF measurements, higher velocities of up to \SI{8}{\milli\meter\per\second} (dark red) are located at the top of the cross-section inside the sinusoidal corrugation. Thus, the sinusoidal corrugation can be directly correlated with the local velocity increase in the radial direction. Furthermore, the arrows in the magnification show the formation of vortices in the cross-section and support the assumption of mixed flow in the sinusoidal corrugation. As mentioned in Section~\ref{sec:SoluteSimulation}, this mixing is caused by the fluid flowing around the sinusoidal corrugation and thus being redirected permanently. This effect can only be visualized when averaging the MRI signal over a length of \SI{10}{\milli\meter}, which corresponds to several sinus waves. Low fluid-flow velocities down to \SI{0}{\milli\meter\per\second} are indicated in dark blue. No significant radial velocities could be measured in the lower part of the module that corresponds to the membrane and the permeate channels. Inside the permeate channels, the purified solution is transported towards the outlet, and axial (in-plane) velocities predominate (that are not shown in this study).

In the HF measurements (slice thickness of \SI{0.5}{\milli\meter}, roughly a quarter of one sinus wave), high velocities of up to \SI{3}{\milli\meter\per\second} are located inside one half of a sinusoidal corrugation. These velocities mostly point straight upwards or downwards, sometimes even both directions in one single sinus wave. This velocity distribution pattern of fluid flowing upwards and downwards in one single sinus wave is in good agreement with the simulation results of fluid flow in z-direction shown in Figure~\ref{fig:velocityprofile}~d). These upward and downward velocities show strong fluid interaction to and away from the membrane, explaining the better performance of the sinusoidal corrugation.

Only the combination of non-invasive measurements with bigger (low-field NMR imaging) and smaller (high-field NMR imaging) slice thickness reveal and characterize all hydrodynamic effects in small filtration devices completely.

\section{Conclusion}
The comprehensive methodology to design fluid compartments minimizing concentration polarization resulted in a novel high-throughput membrane module optimized for flexible use in ultrafiltration screenings. The investigated built-in structures are suitable for one-step production, reducing manufacturing complexity significantly.

Simulations focusing on mass transfer and permeate-flux modeling visualize that the analyzed built-in structures influence velocity gradients and thus the shear rate distribution at the membrane. They show good agreement between regions of high shear rates and low concentrations on the membrane surface and explain the observed permeate flux enhancement. The results suggest that the cross-sectional narrowing and the alternating structures not only increase shear rates at the membrane but also favor diffusion and thus mass transfer. Both, simulations and experiments prove the highest permeate volume flux for the sinusoidal corrugation. This structure outperforms the conventional net-shaped spacer in experiments with BSA by \SI{10}{\percent} to \SI{30}{\percent} higher permeate fluxes; followed by the herringbone structure and the empty channel. The reasons for the increased permeate flux were confirmed and elucidated using high- and low-field magnetic resonance imaging. Low-field measurements showed a mixing effect when averaging over several sinus waves as fluid flows around each sinus wave. High-field measurements revealed the flow pointing towards and away from the membrane in one single sinus wave. These two overlapping effects favor the filtration performance inside the membrane module. An extensive parameter study could further optimize the sinusoidal geometry. 

Where other studies focus on simplifications and fundamental research of fluid flow, this study handles realistic built-in structures, which can directly be implemented in a commercial membrane module. The proposed CFD simulation can detect concentration accumulation and dead zones in membrane modules in advance and supports module design. The proposed species-retention model is an acceptable first-order approach and elucidates the interaction of hydrodynamics and membrane transport during membrane filtration affected by local hydrodynamic conditions. For a more representative ultrafiltration model, the concentration-dependent properties such as diffusion coefficient, viscosity, and density should be implemented. Future studies could focus on in-situ biofilm formation to further correlate hydrodynamics with shear stress and time-dependent fouling development.

The new module design with built-in structures improves the filtration process due to mass transfer enhancement and makes conventional net-shaped inlays obsolete. Due to the simplicity of the structure and the practical manufacturing step of injection molding, an enlargement of the module to more membrane area would be well conceivable. In the future, spacer structures could also be modified according to specific applications (e.g., high or low viscous concentration profiles, shear sensitive products like cells).

\section*{Acknowledgement}
M.W. acknowledges DFG funding through the Gottfried Wilhelm Leibniz Award 2019 (grant ID = WE 4678/12-1). M.W. acknowledges the support through an Alexander-von-Humboldt Professorship and the European Research Council (ERC) under the European Union’s Horizon 2020 research and innovation program (grant agreement no. 694946). This work was also performed in part at the Center for Chemical Polymer Technology CPT, which is supported by the EU and the federal state of North Rhine-Westphalia (grant no. EFRE 30 00 883 02). Simulations were performed with computing resources granted by RWTH Aachen University. The authors thank Univ.-Prof. Dr. med. Fabian Kiessling and Univ.-Prof. Dr.-Ing. Volkmar Schulz (both: Center for Biohybrid Medical Systems (CBMS), RWTH Aachen University, 52074 Aachen, Germany) for giving us the opportunity to use a high-field tomography system. The authors also thank Dr. Nicolas Groß-Weege and Teresa Nolte, M.Sc., for their support in high-field measurements.

 \section*{References}
\bibliographystyle{elsarticle-num}
\biboptions{sort&compress}
\bibliography{Publication1}{}

\newcommand{\noop}[1]{}
\begin{thebibliography}{10}
\expandafter\ifx\csname url\endcsname\relax
  \def\url#1{\texttt{#1}}\fi
\expandafter\ifx\csname urlprefix\endcsname\relax\def\urlprefix{URL }\fi
\expandafter\ifx\csname href\endcsname\relax
  \def\href#1#2{#2} \def\path#1{#1}\fi

\bibitem{Belfort1994}
G.~Belfort, R.~H. Davis, A.~L. Zydney, The behavior of suspensions and
  macromolecular solutions in crossflow microfiltration, Journal of Membrane
  Science 96~(1) (1994) 1 -- 58.

\bibitem{VanReis2007}
R.~van Reis, A.~Zydney, {Bioprocess membrane technology}, Journal of Membrane
  Science 297~(1-2) (2007) 16--50.

\bibitem{Charcosset2012}
C.~Charcosset, Membrane Processes in Biotechnology and Pharmaceutics, {Elsevier
  Science}, Burlington, 2012.

\bibitem{ALAANI.2020}
S.~{Al Aani}, T.~N. Mustafa, N.~Hilal, Ultrafiltration membranes for wastewater
  and water process engineering: A comprehensive statistical review over the
  past decade, {Journal of Water Process Engineering} 35 (2020) 101241.

\bibitem{Lutz2015}
H.~Lutz (Ed.), Ultrafiltration for bioprocessing: Development and
  implementation or robusts processes, Vol.~29 of Woodhead Publishing series in
  biomedicine, {Woodhead Publishing}, Cambridge, 2015.

\bibitem{Belfort1989}
G.~Belfort, Fluid mechanics in membrane filtration: Recent developments,
  Journal of Membrane Science 40~(2) (1989) 123--147.

\bibitem{vandenBerg1990}
{G. B. van den Berg, C. A. Smolders}, Flux decline in ultrafiltration
  processes, Desalination 77 (1990) 101--133.

\bibitem{Matthiasson1980}
E.~Matthiasson, B.~Sivik, Concentration polarization and fouling, Desalination
  35 (1980) 59 -- 103.

\bibitem{Bhattacharjee2020}
C.~Bhattacharjee, V.~K. Saxena, S.~Dutta, Static turbulence promoters in
  cross-flow membrane filtration: a review, Chemical Engineering Communications
  207~(3) (2020) 413--433.

\bibitem{Geraldes2002}
V.~Geraldes, V.~Semiao, M.~Norberta~de Pinho, Flow management in nanofiltration
  spiral wound modules with ladder-type spacers, Journal of Membrane Science
  203 (2002) 87--102.

\bibitem{Rodrigues2012}
C.~Rodrigues, V.~Geraldes, M.~N. de~Pinho, V.~Semi{\~a}o, Mass-transfer
  entrance effects in narrow rectangular channels with ribbed walls or
  mesh-type spacers, Chemical Engineering Science 78 (2012) 38--45.

\bibitem{Qamar2019}
A.~Qamar, S.~Bucs, C.~Picioreanu, J.~Vrouwenvelder, N.~Ghaffour, Hydrodynamic
  flow transition dynamics in a spacer filled filtration channel using direct
  numerical simulation, Journal of Membrane Science 590 (2019) 117264.

\bibitem{Schwinge2002}
J.~Schwinge, D.~E. Wiley, D.~F. Fletcher, Simulation of the flow around spacer
  filaments between narrow channel walls. 1. hydrodynamics, Industrial {\&}
  Engineering Chemistry Research 41~(12) (2002) 2977--2987.

\bibitem{DaCosta1993}
{A.R. Da Costa, A.G. Fane, D.E. Wiley}, Ultrafiltration of whey protein
  solutions in spacer-filled flat channels, Journal of Membrane Science
  76~(1-3) (1993) 245--254.

\bibitem{Picioreanu2009}
C.~Picioreanu, J.~S. Vrouwenvelder, M.~{van Loosdrecht}, Three-dimensional
  modeling of biofouling and fluid dynamics in feed spacer channels of membrane
  devices, Journal of Membrane Science 345~(1-2) (2009) 340--354.

\bibitem{Karode2001}
S.~K. Karode, A.~Kumar, {Flow visualization through spacer filled channels by
  computational fluid dynamics I.: Pressure drop and shear rate calculations
  for flat sheet geometry}, Journal of Membrane Science 193~(1) (2001) 69 --
  84.

\bibitem{Li2002}
F.~Li, W.~Meindersma, A.~B. de~Haan, T.~Reith, Optimization of commercial net
  spacers in spiral wound membrane modules, Journal of Membrane Science
  208~(1-2) (2002) 289--302.

\bibitem{DaCosta1991}
{A. R. Da Costa and A. G. Fane and C. J. D. Fell and A. C. M. Franken}, Optimal
  channel spacer design for ultrafiltration, Journal of Membrane Science 62~(3)
  (1991) 275 -- 291.

\bibitem{Kavianipour2017}
O.~Kavianipour, G.~D. Ingram, H.~B. Vuthaluru, Investigation into the
  effectiveness of feed spacer configurations for reverse osmosis membrane
  modules using computational fluid dynamics, Journal of Membrane Science 526
  (2017) 156--171.

\bibitem{Abid2017}
H.~S. Abid, D.~J. Johnson, R.~Hashaikeh, N.~Hilal, A review of efforts to
  reduce membrane fouling by control of feed spacer characteristics,
  Desalination 420 (2017) 384--402.

\bibitem{Lee2016}
J.-Y. Lee, W.~S. Tan, J.~An, C.~K. Chua, C.~Y. Tang, A.~G. Fane, T.~H. Chong,
  {The potential to enhance membrane module design with 3D printing
  technology}, Journal of Membrane Science 499 (2016) 480--490.

\bibitem{FimbresWeihs2010}
G.~A. Fimbres-Weihs, D.~E. Wiley, {Review of 3D CFD modeling of flow and mass
  transfer in narrow spacer-filled channels in membrane modules}, Chemical
  Engineering and Processing: Process Intensification 49~(7) (2010) 759--781.

\bibitem{Koo2021}
J.~W. Koo, J.~S. Ho, J.~An, Y.~Zhang, C.~K. Chua, T.~H. Chong, A review on
  spacers and membranes: Conventional or hybrid additive manufacturing?, Water
  research 188 (2021) 116497.

\bibitem{Kodym2011}
R.~Kod{\'y}m, F.~Vlas{\'a}k, D.~{\v{S}}nita, A.~{\v{C}}ern{\'i}n, K.~Bouzek,
  Spatially two-dimensional mathematical model of the flow hydrodynamics in a
  channel filled with a net-like spacer, Journal of Membrane Science 368~(1-2)
  (2011) 171--183.

\bibitem{Rivera2017}
F.~F. Rivera, L.~Fabi{\'a}n, P.~Hidalgo, G.~Orozco, {Study of Hydrodynamics at
  Asahi\texttrademark~prototype electrochemical flow reactor, using
  computational fluid dynamics and experimental characterization techniques},
  Electrochimica Acta 245 (2017) 107--117.

\bibitem{Bucs2015}
S.~S. Bucs, R.~V. Linares, J.~O. Marston, A.~I. Radu, J.~S. Vrouwenvelder,
  C.~Picioreanu, Experimental and numerical characterization of the water flow
  in spacer-filled channels of spiral-wound membranes, Water Research 87 (2015)
  299 -- 310.

\bibitem{Thiess2017}
H.~Thiess, M.~Leuthold, U.~Grummert, J.~Strube, Module design for
  ultrafiltration in biotechnology: Hydraulic analysis and statistical
  modeling, Journal of Membrane Science 540 (2017) 440--453.

\bibitem{Johannink2015}
M.~Johannink, K.~Masilamani, A.~Mhamdi, S.~Roller, W.~Marquardt, Predictive
  pressure drop models for membrane channels with non-woven and woven spacers,
  Desalination 376 (2015) 41--54.

\bibitem{Ahmad2005}
A.~L. Ahmad, K.~K. Lau, M.~Z. {Abu Bakar}, Impact of different spacer filament
  geometries on concentration polarization control in narrow membrane channel,
  Journal of Membrane Science 262~(1-2) (2005) 138--152.

\bibitem{Cao2001}
{Z. Cao, D.E. Wiley, A.G. Fane}, {CFD simulations of net-type turbulence
  promoters in a narrow channel}, Journal of Membrane Science 185~(2) (2001)
  157--176.

\bibitem{Balster2010}
J.~Balster, D.~F. Stamatialis, M.~Wessling, {Membrane with integrated spacer},
  Journal of Membrane Science 360~(1-2) (2010) 185--189.

\bibitem{Racz1986}
I.~G. R{\'a}cz, J.~Wassink, R.~Klaassen, Mass transfer, fluid flow and membrane
  properties in flat and corrugated plate hyperfiltration modules, Desalination
  60~(3) (1986) 213--222.

\bibitem{Mazinani.2019}
S.~Mazinani, A.~Al-Shimmery, Y.~J. Chew, D.~Mattia, {3D Printed
  Fouling-Resistant Composite Membranes}, ACS Applied Materials \& Interfaces
  11~(29) (2019) 26373--26383.

\bibitem{Lee2013}
Y.~K. Lee, Y.-J. Won, J.~H. Yoo, K.~H. Ahn, C.-H. Lee, Flow analysis and
  fouling on the patterned membrane surface, Journal of Membrane Science 427
  (2013) 320--325.

\bibitem{Zhou2021}
Z.~Zhou, B.~Ling, I.~Battiato, S.~M. Husson, D.~A. Ladner, Concentration
  polarization over reverse osmosis membranes with engineered surface features,
  {Journal of Membrane Science} 617 (2021) 118199.

\bibitem{LiraTeco2016}
J.~E. Lira-Teco, F.~Rivera, O.~Far{\'i}as-Moguel, J.~Torres-Gonz{\'a}lez,
  Y.~Reyes, R.~Anta{\~n}o-L{\'o}pez, G.~Orozco, F.~Casta{\~n}eda-Zaldivar,
  {Comparison of experimental and CFD mass transfer coefficient of three
  commercial turbulence promoters}, Fuel 167 (2016) 337--346.

\bibitem{Completo2016}
C.~Completo, V.~Semiao, V.~Geraldes, {Efficient CFD-based method for designing
  cross-flow nanofiltration small devices}, Journal of Membrane Science 500
  (2016) 190--202.

\bibitem{Santos2007}
J.~Santos, V.~Geraldes, S.~Velizarov, J.~G. Crespo, {Investigation of flow
  patterns and mass transfer in membrane module channels filled with
  flow-aligned spacers using computational fluid dynamics (CFD)}, Journal of
  Membrane Science 305~(1-2) (2007) 103--117.

\bibitem{Liu2013}
J.~Liu, A.~Iranshahi, Y.~Lou, G.~Lipscomb, Static mixing spacers for spiral
  wound modules, Journal of Membrane Science 442 (2013) 140--148.

\bibitem{Schwinge2000}
{J. Schwinge, D. E. Wiley, A. G. Fane, R. Guenther}, Characterization of a
  zigzag spacer for ultrafiltration, Journal of Membrane Science 172~(1-2)
  (2000) 19--31.

\bibitem{Jung2019}
S.~Y. Jung, J.~E. Park, T.~G. Kang, K.~H. Ahn, {Design Optimization for a
  Microfluidic Crossflow Filtration System Incorporating a Micromixer},
  {Micromachines} 10~(12) (2019) 836.

\bibitem{Shrivastava2008}
A.~Shrivastava, S.~Kumar, E.~L. Cussler, Predicting the effect of membrane
  spacers on mass transfer, Journal of Membrane Science 323~(2) (2008)
  247--256.

\bibitem{Schwinge2002b}
J.~Schwinge, D.~E. Wiley, D.~F. Fletcher, Simulation of the flow around spacer
  filaments between channel walls. 2. mass-transfer enhancement, Industrial
  {\&} Engineering Chemistry Research 41~(19) (2002) 4879--4888.

\bibitem{Koutsou2004}
C.~P. Koutsou, S.~G. Yiantsios, A.~J. Karabelas, Numerical simulation of the
  flow in a plane-channel containing a periodic array of cylindrical turbulence
  promoters, Journal of Membrane Science 231~(1-2) (2004) 81--90.

\bibitem{Schock1987}
G.~Schock, A.~Miquel, Mass transfer and pressure loss in spiral wound modules,
  Desalination 64 (1987) 339 -- 352.

\bibitem{vanderWaal1989}
M.~van~der Waal, S.~Stevanovic, I.~Racz, {Mass transfer in corrugated-plate
  membrane modules. II. Ultrafiltration experiments}, Journal of Membrane
  Science 40~(2) (1989) 261--275.

\bibitem{BALSTER2006}
J.~Balster, I.~P{\"u}nt, D.~Stamatialis, M.~Wessling, Multi-layer spacer
  geometries with improved mass transport, Journal of Membrane Science
  282~(1-2) (2006) 351 -- 361.

\bibitem{Wiese.2018}
M.~Wiese, S.~Benders, B.~Bl{\"u}mich, M.~Wessling, {3D MRI velocimetry of
  non-transparent 3D-printed staggered herringbone mixers}, {Chemical
  Engineering Journal} 343 (2018) 54--60.

\bibitem{FRITZMANN.2013}
C.~Fritzmann, M.~Hausmann, M.~Wiese, M.~Wessling, T.~Melin, Microstructured
  spacers for submerged membrane filtration systems, Journal of Membrane
  Science 446 (2013) 189 -- 200.

\bibitem{FRITZMANN.2014}
C.~Fritzmann, M.~Wiese, T.~Melin, M.~Wessling, Helically microstructured
  spacers improve mass transfer and fractionation selectivity in
  ultrafiltration, Journal of Membrane Science 463 (2014) 41 -- 48.

\bibitem{ARMBRUSTER.2018}
S.~Armbruster, O.~Cheong, J.~L{\"o}lsberg, S.~Popovic, S.~Y{\"u}ce,
  M.~Wessling, {Fouling mitigation in tubular membranes by 3D-printed
  turbulence promoters}, Journal of Membrane Science 554 (2018) 156 -- 163.

\bibitem{Popovic2015}
S.~Popovi\'{c}, M.~Wessling, Turbulence Promoters in Membrane Processes,
  Journal of Membrane Science Virtual Issue, 2015.

\bibitem{Femmer2016f}
T.~Femmer, I.~Flack, M.~Wessling, {Additive Manufacturing in Fluid Process
  Engineering}, Chemie Ingenieur Technik 88~(5) (2016) 535--552.

\bibitem{Gibson2015}
I.~Gibson, D.~Rosen, B.~Stucker, Additive manufacturing technologies: 3D
  printing, rapid prototyping and direct digital manufacturing, {2nd} Edition,
  Springer, 2015.

\bibitem{AlObaidi2016}
M.~A. Al-Obaidi, I.~M. Mujtaba, Steady state and dynamic modeling of spiral
  wound wastewater reverse osmosis process, Computers {\&} Chemical Engineering
  90 (2016) 278--299.

\bibitem{Sundaramoorthy2011}
S.~Sundaramoorthy, G.~Srinivasan, D.~Murthy, {An analytical model for spiral
  wound reverse osmosis membrane modules: Part II --- Experimental validation},
  Desalination 277~(1-3) (2011) 257--264.

\bibitem{Wypysek2019}
D.~Wypysek, D.~Rall, M.~Wiese, T.~Neef, G.-H. Koops, M.~Wessling, {Shell and
  lumen side flow and pressure communication during permeation and filtration
  in a multibore polymer membrane module}, {Journal of Membrane Science}
  584~(8) (2019) 254--267.

\bibitem{Luelf.2018}
T.~Luelf, D.~Rall, D.~Wypysek, M.~Wiese, T.~Femmer, C.~Bremer, J.~U. Michaelis,
  M.~Wessling, {3D-printed rotating spinnerets create membranes with a twist},
  {Journal of Membrane Science} 555 (2018) 7--19.

\bibitem{Wiese.2018b}
M.~Wiese, C.~Malkomes, B.~Krause, M.~Wessling, {Flow and filtration imaging of
  single use sterile membrane filters}, {Journal of Membrane Science} 552
  (2018) 274--285.

\bibitem{Wiese.2019}
M.~Wiese, O.~Nir, D.~Wypysek, L.~Pokern, M.~Wessling, {Fouling minimization at
  membranes having a 3D surface topology with microgels as soft model
  colloids}, {Journal of Membrane Science} 569 (2019) 7--16.

\bibitem{WYPYSEK.2020}
D.~Wypysek, A.~M. Kalde, F.~Pradellok, M.~Wessling, In-situ investigation of
  wetting patterns in polymeric multibore membranes via magnetic resonance
  imaging, {Journal of Membrane Science} (2020) 119026.

\bibitem{DaCosta1994}
A.~{Da Costa}, A.~Fane, D.~Wiley, {Spacer characterization and pressure drop
  modelling in spacer-filled channels for ultrafiltration}, Journal of Membrane
  Science 87~(1) (1994) 79--98.

\bibitem{Chew2004}
J.~Chew, W.~Paterson, I.~Wilson, Fluid dynamic gauging for measuring the
  strength of soft deposits, Journal of Food Engineering 65 (2004) 175--187.

\bibitem{Haidari2016}
A.~Haidari, S.~Heijman, W.~{van der Meer}, {Visualization of hydraulic
  conditions inside the feed channel of Reverse Osmosis: A practical comparison
  of velocity between empty and spacer-filled channel}, Water Research 106
  (2016) 232--241.

\bibitem{Koutsou2015}
C.~P. Koutsou, A.~J. Karabelas, {A novel retentate spacer geometry for improved
  spiral wound membrane (SWM) module performance}, Journal of Membrane Science
  488 (2015) 129--142.

\bibitem{Vrouwenvelder2009}
J.~Vrouwenvelder, D.~{Graf von der Schulenburg}, J.~Kruithof, M.~Johns, M.~{van
  Loosdrecht}, {Biofouling of spiral-wound nanofiltration and reverse osmosis
  membranes: A feed spacer problem}, Water Research 43~(3) (2009) 583--594.

\end{thebibliography}



\end{document}